\documentclass[graybox]{svmult}

\usepackage{mathptmx}       
\usepackage{helvet}         
\usepackage{courier}        
\usepackage{type1cm}        
\usepackage{makeidx}         
\usepackage{graphicx}        
\usepackage{multicol}        
\usepackage[bottom]{footmisc}
\usepackage{amssymb}

\def\R{{\mathbb{R}}}

\makeindex             

\begin{document}

\title*{Dynamic trajectory control  of gliders}
\author{Rui Dil\~ao\inst{1,2} \and Jo\~ao Fonseca\inst{1}}
\institute{Nonlinear Dynamics Group, Instituto Superior
T\'ecnico, Av. Rovisco Pais, 1049-001 Lisbon, Portugal \   \texttt{rui@sd.ist.utl.pt},  \texttt{jlpfonseca@gmail.com} \and Institut des Hautes \'Etudes Scientifiques, 35, route de Chartres, 91440 Bures-sur-Yvette, France.}
%
%
\maketitle

\abstract*{A new  dynamic control algorithm in order to direct the trajectory of  a glider to a pre-assigned target point is proposed.
The algorithms runs iteratively and the approach to the target point is self-correcting. The algorithm is applicable to any non-powered lift-enabled vehicle (glider) travelling in planetary atmospheres. As a proof of concept, we have 
applied the new algorithm to the command and control of the trajectory of the Space Shuttle during the Terminal Area Energy Management (TAEM) phase.}

\abstract{A new  dynamic control algorithm in order to direct the trajectory of  a glider to a pre-assigned target point is proposed.
The algorithms runs iteratively and the approach to the target point is self-correcting. The algorithm is applicable to any non-powered lift-enabled vehicle (glider) travelling in planetary atmospheres. As a proof of concept, we have 
applied the new algorithm to the command and control of the trajectory of the Space Shuttle during the Terminal Area Energy Management (TAEM) phase.
}

\section{Introduction}
\label{sec:1}
Space vehicles travel at extreme conditions of speed and acceleration that typically do not allow for a ``man-in-the loop" approach, forcing, at least partially, automation of the flight controls. 
Thus, automated guidance and control systems are a critical component for any re-usable  space flight vehicle.

For example, the implementation of control mechanisms for atmosphere re-entry and automatic landing  systems used in the Space Shutle  focused either on pre-programmed manoeuvres following a nominal pre-computed trajectory, or hopping across different nominal trajectories whenever the vehicle deviates from an initially selected trajectory, \cite{3} and \cite{4}.

A typical return flight from space  has three main phases:
\begin{description}
\item{1)}	Atmospheric re-entry phase:  In this initial re-entry phase the transition from spacecraft to aircraft flight mode occurs. The typical altitudes for this phase are in the range 120-40~km.
\item{2)}	Glide to the landing site phase,   usually   referred as Terminal Area Energy Management (TAEM),  occurring  in the altitude range   40-3~km.
\item{3)}	Final approach and landing phase, occurring  in the altitude range  3-0~km.
\end{description}

While in the atmosphere re-entry phase, the biggest priority is to ensure that the structural constraints of the vehicle are not exceeded; during the TAEM phase, the biggest priority is to ensure that the vehicle reaches the  Heading Alignment Circle  (HAC) where preparation for landing is initiated.

On a typical mission, the TAEM phase begins at the altitude of $25,000-40,000$~m  at a speed around $2 - 6$~M (Mach), and  
finishes at  the HAC at the altitude  of $1,500- 3,000$ ~m, with a speed of the order of $0.20$~M.

In this paper, we propose a new  dynamic control algorithm in order to redirect the trajectory of  gliders to a pre-assigned target point.
This algorithm runs iteratively enabling a self-correcting approach to the HAC and is applicable to any  non-powered lift-enabled vehicle (glider) travelling in planetary atmospheres.

This paper is organised as follows. In section~\ref{sec:2}, we present the equations of motion of a glider and we discuss the approximations we use to define the controllability conditions.  In section~\ref{sec:3}, we briefly discuss the phenomenology of aircraft gliding motion,  instrumental for the design of a dynamic control strategy. In section~\ref{sec:4}, we derive the dynamic control   algorithm,  and in section~\ref{sec:5} we present realistic simulations for the Space Shuttle TAEM guidance and control. Finally, in section~\ref{sec:6} we discuss the main conclusions of the paper.

\section{Gliding motion}
\label{sec:2}

\begin{figure}
\begin{centering}
 \includegraphics[width= 0.6\hsize]{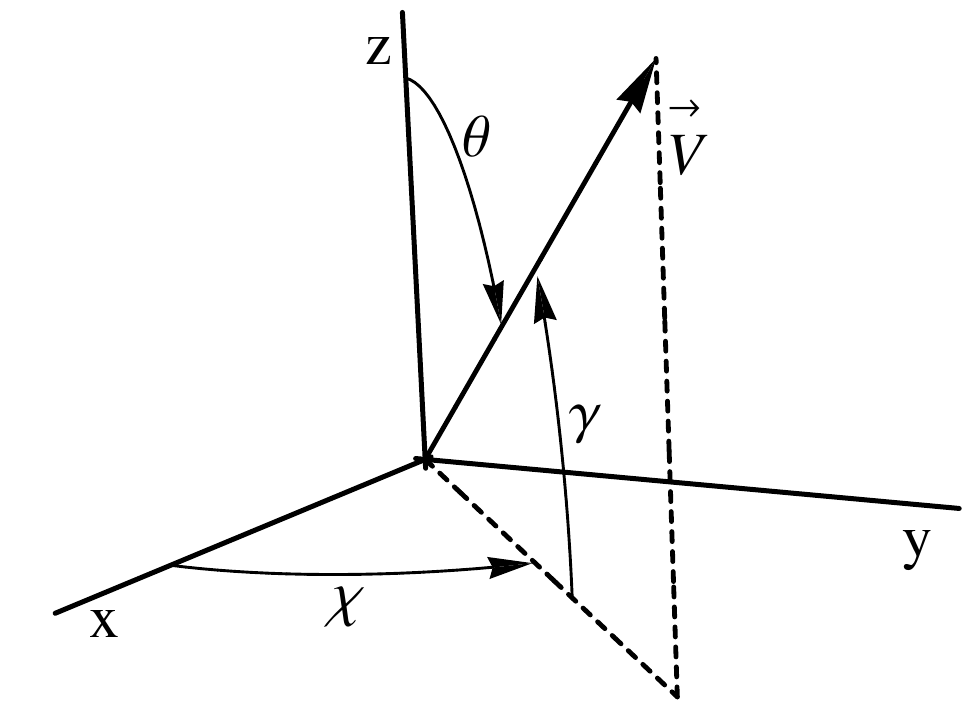}
 \caption{Local coordinate system for the point mass glider model. The origin of coordinates is located at the centre of mass of the aircraft, and the vector ${\vec V}$ is the velocity vector not necessarily collinear with the aircraft longitudinal axis.} 
\label{fig1} 
\end{centering}
\end{figure}

\begin{figure}
\begin{centering}
 \includegraphics[width= 0.9\hsize]{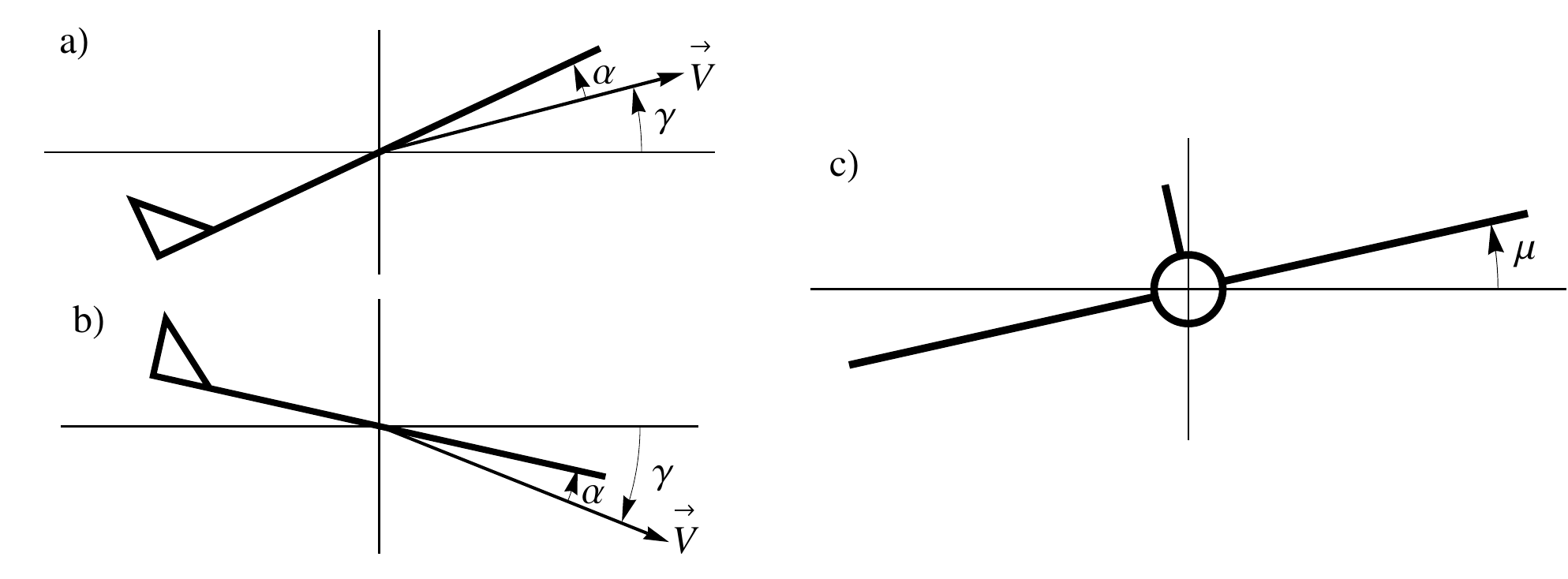}
 \caption{In a) and b), we show the flight path angle $\gamma$ and the angle of attack $\alpha$ of an aircraft. In c) we show the bank angle $\mu$, measuring the inclination of the aircraft in the plane containing the velocity vector and the horizontal direction.  The flight path angle $\gamma$ depends on the angle of attack, on the aerodynamic coefficients of the aircraft and on the  Mach number. The control of a  glider is done by the manipulation of the angles of attack and  bank.}
\label{fig2} 
\end{centering}
\end{figure}

We consider that aircraft gliding motion in a planetary atmosphere is well described by a point mass vehicle model under the influence of a gravity field, \cite{7},  \cite{8}, \cite{9} and \cite{10}. In this case, the equations of motion of a gliding aircraft (no thrust forces) are,
\begin{equation}
\left\{\begin{array}{l}
m\dot{V}=-mg(z) \sin{\gamma}-D(\alpha, Ma) \\
mV\dot{\gamma}=-mg(z) \cos{\gamma}+L(\alpha, Ma)\cos{\mu} \\\
mV\dot{\chi}\cos{\gamma}=L\sin{\mu}  \
\end{array}\right. \, ,
\left\{\begin{array}{l}
\dot x=V \cos{\chi}\cos{\gamma}\\
\dot y=V \sin{\chi}\cos{\gamma}\\
\dot z=V \sin{\gamma}
\end{array}\right.
\label{equations}
\end{equation}
where $m$ is the aircraft mass, $V=\sqrt{V_x^2+V_y^2+V_z^2}$ is the aircraft speed,  $\gamma $ is the flight path angle  as defined in figures~\ref{fig1} and \ref{fig2},  $\mu$ is the bank angle as defined in figure~\ref{fig2}c), $D(\alpha, Ma)$ and $L(\alpha, Ma)$ are the drag and lift forces induced by the atmosphere, $\alpha$ is the angle of attack and $Ma$ is the Mach number.  In general,   the Mach number $Ma$ is a function of $V$ and $z$.  
The function 
$g(z)=g_0(R_E/(R_E+z))^2$ is the gravity acceleration, $g_0=9.80665$~m/s$^2$ is the Earth standard gravitational acceleration constant  and $R_E=6.371\times10^6$~m is the Earth (or planetary) mean radius.

In the local reference frame of the aircraft, figure~\ref{fig1},  $V\in(0,\infty)$, $\gamma \in [-\pi/2,\pi/2]$ and $\chi \in [0,2\pi]$. The bank angle $\mu$ is defined in the interval $ [-\pi/2,\pi/2]$. In this reference frame,  positive values of $\mu$ correspond to left turns and negative values of $\mu$ correspond to right turns. As usual, $(x,y,z)\in \R^3$ and $(\dot x,\dot y,\dot z)\in \R^3$.
 In the system of equations (\ref{equations}), $\alpha$ and $\mu$ can be seen as input parameters. 
 
 To define the local system of coordinates, we have used a flat-Earth approach.  As we want to analyse the motion of gliders during the TAEM phase,  the height at which the TAEM phase starts is very small when compared to the Earth radius, justifying our analysis. However, this approach can be further refined by using an ellipsoidal coordinate system adequate to EarthÕs shape, such as   the WGS-84 coordinate system.

In figure~\ref{fig2}a)-b), we show the angle of attack $\alpha$ defined as the angle between the longitudinal reference line of the aircraft and the vector velocity of the aircraft. 
In  airplanes, the angle of attack is always a positive angle. 
While in most aircrafts attack angles are always smaller than $15^o$, the Space Shuttle is capable of attack angles up to $45^o$, \cite{5} and \cite{6}. In figure~\ref{fig2}c), we show the bank angle, defining the inclination of the aircraft in the plane containing the velocity vector.

The drag and lift forces in the system of equations (\ref{equations}) are given by,
\begin{equation}
\begin{array}{l}
D(\alpha,Ma)=\bar{q}SC_D(\alpha,Ma)= \frac{1}{2}\rho(z)V^2  SC_D(\alpha,Ma)\\[10pt]
L(\alpha,Ma)=\bar{q}SC_L(\alpha,Ma)=\frac{1}{2}\rho(z)V^2  SC_L(\alpha,Ma)
\end{array}
\label{draglift}
\end{equation}
where $\bar{q}=\rho(z)V^2/2$ is the dynamic pressure, $S$ is the wing area of the aircraft, $\rho(z)$ is the atmosphere density as a function of altitude (Appendix) and $Ma$ is the Mach number. For each specific aircraft, the functions $C_D(\alpha,Ma)$ and $C_L(\alpha,Ma)$ are the aerodynamic drag and lift coefficients  determined in wind tunnel experiments. 

Introducing the expressions (\ref{draglift}) into equations (\ref{equations}), we obtain the final form for the equations of motion of a glider,
\begin{equation}
\left\{\begin{array}{l}\displaystyle
\dot{V}=-g(z) \sin{\gamma}-\left(\frac{1}{2 m}\rho(z) SC_D(\alpha,Ma) \right) V^2 \\[10pt]\displaystyle
\dot{\gamma}=-{g(z)\over V} \cos{\gamma}+\left(\frac{1}{2 m}\rho(z) SC_L(\alpha,Ma) \right) V\cos{\mu}  \\ [10pt]\displaystyle
\dot{\chi}=\left(\frac{1}{2 m}\rho SC_L \right) V{\sin{\mu}\over \cos{\gamma}} 
\end{array}\right.   \, ,
\left\{\begin{array}{l}\displaystyle
\dot x=V \cos{\chi}\cos{\gamma}\\[10pt]\displaystyle
\dot y=V \sin{\chi}\cos{\gamma}\\[10pt]\displaystyle
\dot z=V \sin{\gamma}\, .
\end{array}\right.
\label{equations2}
\end{equation}
where $\rho(z)$ is calculated  in the Appendix. 

The aircraft gliding  trajectory is described by the system of equations (\ref{equations2}),  enabling a simple geometric solution of the gliding aircraft control problem.

When a glider is falling under a   gravity field it converges to a  steady state  motion with a constant velocity and constant flight path angle given by, \cite{11},
\begin{equation}
\begin{array}{l}\displaystyle
V^*=\sqrt{\frac{ 2 mg}{ \rho S}}{1\over (C_D^2+C_L^2\cos^2{\mu})^{1/4}}\\ \displaystyle
\gamma^* = -\arctan \frac{C_D}{C_L\cos{\mu}} 
\end{array} \, .
\label{fixed}
\end{equation}
The geometry of the solutions of equation (\ref{equations2}) in phase space are analysed in detail in \cite{11}.

\section{Phenomenology of Space Shuttle gliding motion}\label{sec:3}

Using  wind tunnel data for  the operational range of aircrafts during the TAEM phase,  we have done  fits for the aerodynamic drag and lift coefficients $C_D$ and $C_L$ of the Space Shuttle and these are well described  by the parameterised functions,
\begin{equation}
\begin{array}{l}
C_{L}(\alpha,Ma)=(a_1+a_2 \alpha+a_3  \alpha^2 )  K(Ma)^{b_1+\alpha b_2}\\
C_{D}(\alpha,Ma)=(0.01+f_1 Ma^{f_2}+d_3 \alpha^2 ) K(Ma)^{e_1+\alpha e_2}
\end{array}
\label{c1}
\end{equation}
where,
\begin{equation}
K(Ma)=\frac{1}{2}\left(1+\sqrt{\left|1-\left(\frac{Ma}{M_c}\right)^2\right|}\right)
\label{c21}
\end{equation}
is a simplification of the Van Karman functions  expanded to supersonic regimes, \cite{5}. In  table~\ref{tab1}, we show, for the Space Shuttle, the parameter estimation of expressions (\ref{c1}) and (\ref{c21}) with wind tunnel data.

\begin{table}[h!]
\begin{center}
\begin{tabular}{|c|c|c|c|c|}
\hline
Parameter & Estimated & Standard error & t-statistics & P-value \\
\hline
$a_1$ & $-0.053$  & $0.009$ & $-6.15$ & $9.8\times 10^{-8}$  \\
\hline
$a_2$ & $2.73$ &  $0.06$ &  $43.0$ & $1.8\times 10^{-43}$   \\
\hline
$a_3$ & $-1.55$ &  $0.09$ &  $-18.0$ & $2.0\times 10^{-24}$   \\
\hline
$b_1$ & $-1.01$ &  $0.09$ &  $-11.3$ & $7.4\times 10^{-16}$   \\
\hline
$b_2$ & $1.1$ &  $0.1$ &  $8.7$ & $7.6\times 10^{-12}$   \\
\hline
$d_3$ & $1.79$ &  $0.02$ &  $99.0$ & $1.1\times 10^{-63}$   \\
\hline
$e_1$ & $-1.4$ &  $0.1$ &  $-12.6$ & $1.2\times 10^{-17}$   \\
\hline
$e_2$ & $1.5$ &  $0.1$ &  $11.3$ & $5.8\times 10^{-16}$   \\
\hline
$f_1$ & $0.028$ &  $0.004$ &  $6.46$ & $2.9\times 10^{-8}$   \\
\hline
$f_2$ & $1.4$ &  $0.2$ &  $8.57$ & $1.0\times 10^{-11}$   \\
\hline
$M_c$ & $1.25$ &  $0.03$ &  $49.6$ & $1.0\times 10^{-46}$   \\
\hline
\end{tabular}
\end{center}
\caption{Parameters of the aerodynamic drag and lift coefficients (\ref{c1}) for the Space Shuttle, estimated from wind tunnel data, \cite{2}.  The significance of the fits have been determined with a chi-squared test. The large values of the absolute value of the t-statistics measures the likelihood of the parameters in the fits.   The low values of the p-values mean that the fits are highly significant and the probability of finding a value outside the fitted ones are in the range $10^{-8}-10^{-63}$.}
\label{tab1}
\end{table}

Introducing the expressions of $C_{L}(\alpha,Ma)$ and $C_{D}(\alpha,Ma)$ into (\ref{fixed}), changing the angle of attack $\alpha$ and the bank angle $\mu$ leads to changes in the local   steady states of the glider (see (\ref{fixed})), enabling a guided control of the direction of motion and of the glider speed.

To control the  aerodynamic behaviour of an aircraft, two main   parameters are under the control of the aircraft commands: i) the bank angle $\mu$,  and ii) the attack angle $\alpha$.

The bank angle $\mu$ determines the inclination of the aircraft and is used for turn manoeuvres, figure~\ref{fig2}c). 

\begin{figure}[h]
\begin{centering}
 \includegraphics[width= 0.6\hsize]{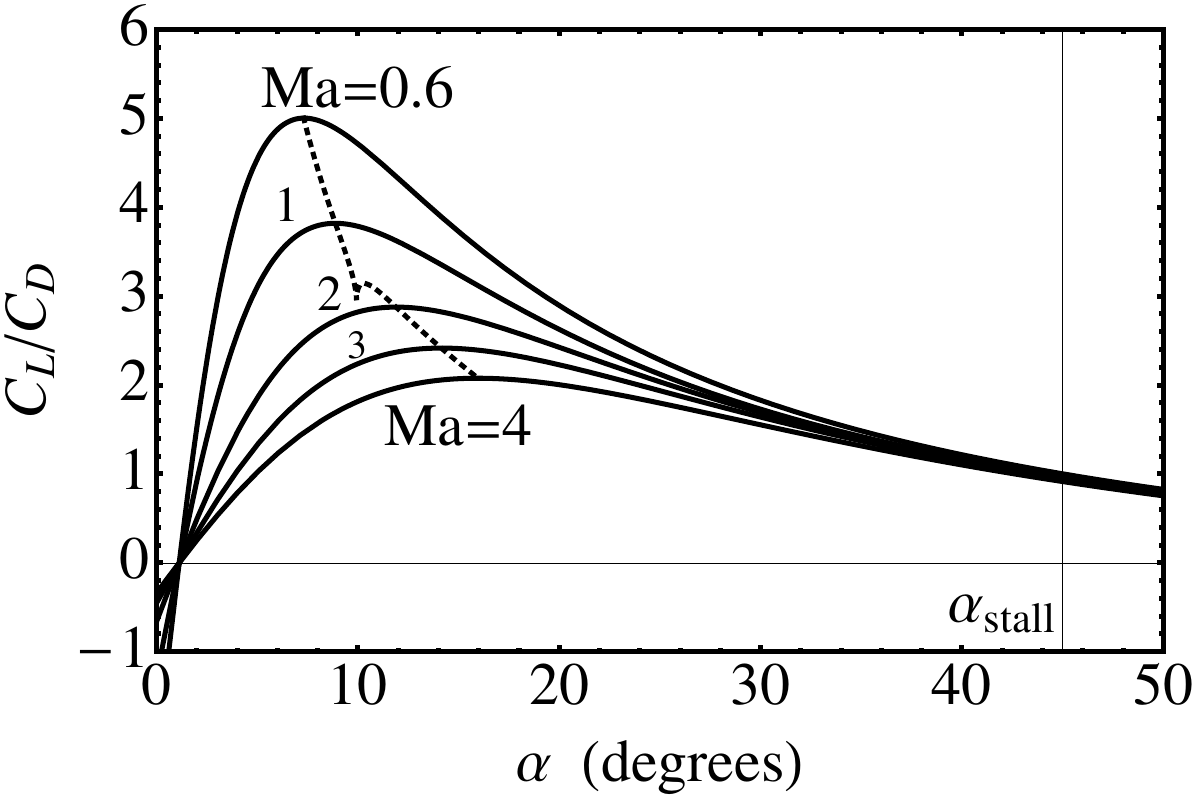}
\caption{Ratio $L/D$, as a function of the angle of attach $\alpha$,  for the Space Shuttle at different Mach numbers, calculated from (\ref{c1})-(\ref{c21}) and table~\ref{tab1}. The no-lift parameter is $\alpha_{nL}=1.5^o$,
 the stall angle is $\alpha_{stall}=45^o$ and $\alpha_{maxgl}$ is given by (\ref{maxgl}). The Space Shuttle is a glider and thus can only move across its $L/D$ curve. For higher speeds this curve will become increasingly flat and the max-glide angle $\alpha_{maxgl}$ will move further to the right reaching saturation.}
\label{fig3} 
\end{centering}
\end{figure}

The no-lift angle $\alpha_{nL}$, the max-glide angle $\alpha_{maxgl}$ and stall angle $\alpha_{stall}$  are particular limits of the angle of attack of an aircraft, figure~\ref{fig2}a) and \ref{fig3}. 

In figure~\ref{fig3}, we show, for several values of the Mach number, the behaviour of the ratio $L/D$, as a function of the angle of attack $\alpha$ for the Space Shuttle. All the curves intersect at the no-lift angle $\alpha_{nL}$. 
The no-lift angle $\alpha_{nL}$  is the angle   for which $L/D$ is zero due the absence of  the lift force and is independent of the speed.
The max-glide angle $\alpha_{maxgl}$ is the angle that maximises the ratio $L/D$, and is dependent on the Mach number.
The stall angle $\alpha_{stall}$ is the angle at which lift dependency looses linearity and lift peaks before beginning to decrease. The stall angle is  independent of the Mach number.

With the  functions (\ref{c1})-(\ref{c21}), we have approximated  the max-glide angle $\alpha_{maxgl}$ as a function of Mach number.
For the case of the Space Shuttle, we have obtained,
\begin{equation}
\alpha_{maxgl}=\left\{
\begin{array}{l}
0.0906+0.0573Ma+0.0071Ma^2\quad (Ma\le 1.25)\\
0.1070+0.0577Ma-0.0037Ma^2\quad (1.25<Ma< 5) 
\end{array}\right.
\label{maxgl}
\end{equation}
determined with a correlation coefficient of  $r^2=0.999$. The Mach number is defined  by $Ma=V/V_{sound}$ where the sound speed is calculated with,
\begin{equation}
V_{sound}=\sqrt{\gamma T(z) R_s}
\label{vsom}
\end{equation}
and $T(z)$ is given in table~\ref{tab2} in the Appendix. $\gamma=1.4$ is the diatomic gas constant and $R_s=287.04$~J/(kg~K).

\section{Dynamic trajectory control of gliders}\label{sec:4}

A glider is not always in an equilibrium state but naturally converges to it given enough time. Our algorithm will take advantage of this behaviour by determining the equilibrium conditions needed to reach the target, imposing them on the system and letting it evolve in time.

To define the control problem, we consider the initial condition,
$$
(x_0,y_0,z_0,V_0, \gamma_0,\chi_0)
$$
defining the initial coordinates of the TAEM phase.
Let,
$$
(x_f,y_f,z_f)
$$
be the space coordinates of the target, which coincide with  the central point in the HAC  region.
We consider that the target point is only defined by the spatial coordinates of the HAC, and the direction of the velocity vector is arbitrary. In fact,  this is possible at low altitudes ($3$~km) because the atmosphere is dense enough to allow the glider to preform turns in short distances and the vehicle is   always travelling near the equilibrium speed.

The intermediate coordinates of the glider path are,
$$
(x_i,y_i,z_i,V_i, \gamma_i,\chi_i)
$$
where $i=0,1,\ldots , f$. These intermediate coordinates are evaluated at time intervals $T_{con}$.

In the configuration space $(x,y,z)$, we define the direction vector from the current  position of the glider to the target point as,
\begin{equation}
\vec P_i=(x_f-x_i,y_f-y_i,z_f-z_i)\, .
\label{pv}
\end{equation}

In order  to  direct the aircraft to the target, we  control the attack and bank angles separately.  

In the \textbf{attack angle heading control}, we analyse the glider trajectory in the three dimensional ambient space $(x,y,z)$, and we command the glider trajectory path angle by controlling the angle of attack $\alpha$. 

In the   \textbf{bank angle heading control}, the control procedures will be done in the $(x,y)$ plane by adjusting the bank angle $\mu$.

\begin{figure}
\begin{centering}
 \includegraphics[width= 0.7\hsize]{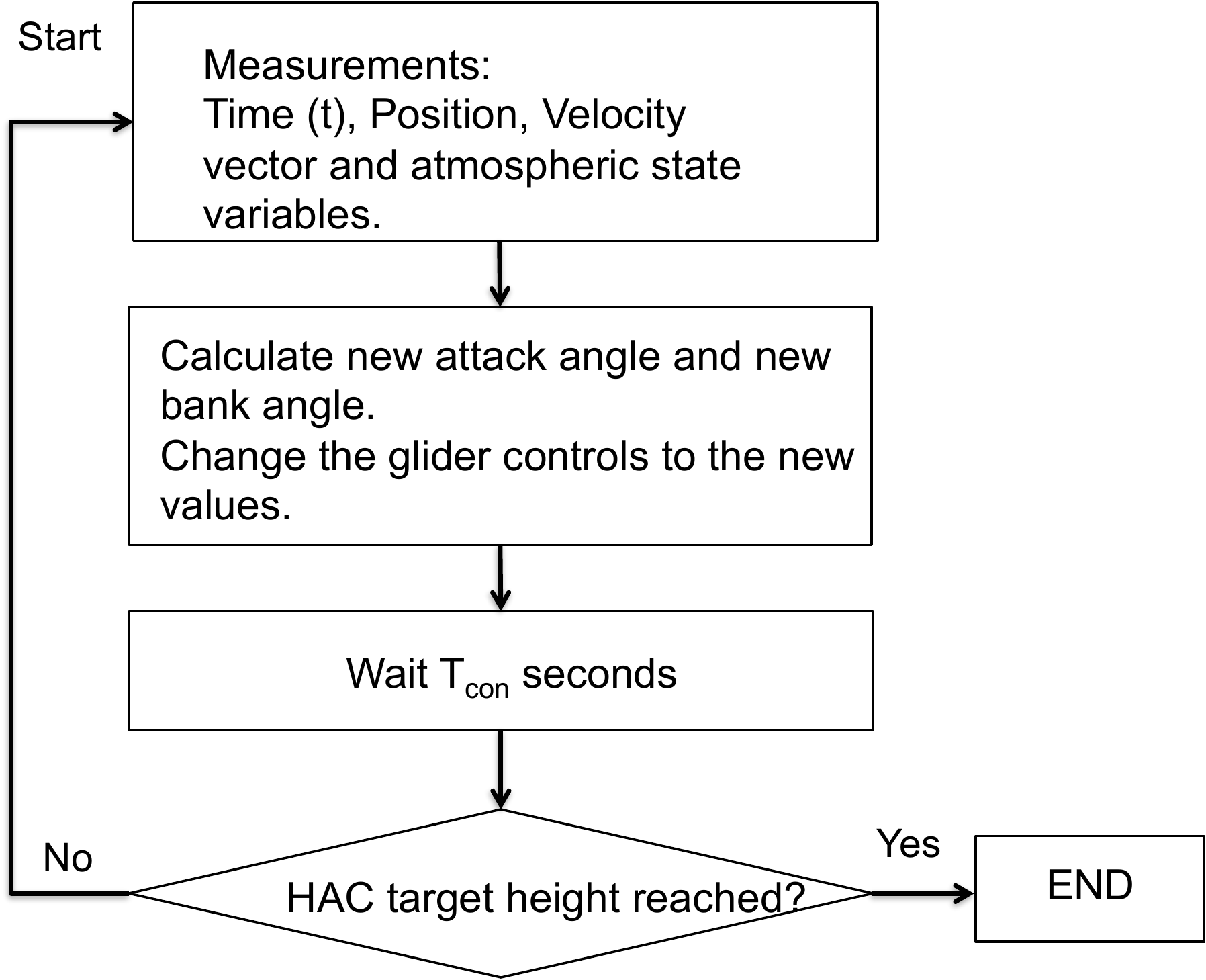}
\caption{Block diagram for the controller algorithm.}
\label{fig4} 
\end{centering}
\end{figure}

At the step number $i$ of the dynamic control process,   the initial conditions  are $(x_i,y_i,z_i,V_i, \gamma_i,\chi_i)$. At this stage, the angle of attack and bank angle are $\alpha_{i}$ and  $\mu_{i}$. Then,  we calculate the new values of the glider control parameters $\alpha_{i+1}$ and  $\mu_{i+1}$ by the two procedures  described below. With these new values for $\alpha$ and $\mu$ , the aircraft will follow a new trajectory  
during  the time interval $T_{con}$, figure~\ref{fig4}.

This control process is  done sequentially in time, until the glider reaches de  HAC region. 
In practical terms, 
the control mechanisms  stops when the distance from the spacecraft to the centre of the HAC point attains a minimum.

We analyse now in detail the two control and command procedures for $\alpha$ and $\mu$.

\begin{description}
\item{\textbf{Attack angle heading control:}} 

The attack angle heading  command and control  was designed so that  the vehicle is always re-orienting vertically  to the HAC point through a straight line path.
 
The tangent of the angle between the $x-y$ projection and the $z$ component of the direction vector $P_i$ to the target point is computed at each iteration, and we obtain,
$$
G_{i+1}=\frac{(z_f-z_i)}{\sqrt{(x_f-x_i)^2+(y_f-y_i)^2}} 
$$
where $(x_i,y_i,z_i)$ is the current position of the glider. At this position, the glider has flight path $\gamma_i$.
Then, to direct the motion of the glider to the target with a steady flight path, by (\ref{fixed}), we must have,
\begin{equation}
G_{i+1}=\tan \gamma=-{1\over C_L/C_{D}\cos{\mu_{i+1}}}\, .
\label{con1}
\end{equation}
Assuming that it is possible to direct the motion to the target using a null bank angle, $\mu_{i+1} =0$, we solve  equation (\ref{con1})  in order to  the ratio $C_L/C_D$, and we obtain the solution $c_{i+1}$.
Then:
\begin{description}
\item{a)} 	If $c_{i+1}$  is bigger than $C_L(\alpha_{maxgl},Ma_i)/C_D(\alpha_{maxgl},Ma_i)$, the target cannot be reached in a straight-line and the max-glide attack angle will be selected, $\alpha_{i+1}=\alpha_{maxgl}$. The curve of $C_L/C_D$ as a function of $\alpha$ and of the Mach number $Ma$ is given by (\ref{c1}) and (\ref{c21}), and $\alpha_{maxgl}$ is calculated from (\ref{maxgl}) and (\ref{vsom}). 

\item{b)} 	If $c_{i+1}$  is smaller than $C_L(\alpha_{stall},Ma_i)/C_D(\alpha_{stall},Ma_i)$, the target cannot be reached in a straight-line and the stall angle will be selected, $\alpha_{i+1}=\alpha_{stall}$.

\item{c)} 	Otherwise, the attack angle  $\alpha_{i+1}$ is computed  by solving the equation $C_L(\alpha,Ma_i)/C_D(\alpha,Ma_i)=c_{i+1}$.
\end{description}

At this stage, we have chosen a new attack angle $\alpha_{i+1}$.
With this new attack angle, we re-orient dynamically and vertically  the aircraft trajectory  to the target. 

\bigskip

\item{\textbf{Bank angle heading control:}}

The bank angle heading  control   was constructed in such a way that, in the $(x,y)$ plan, the aircraft is always re-orienting horizontally to   the HAC. 

The angular misalignment between the direction vector  to the target point (\ref{pv}) and the speed in the $(x,y)$ plane is measured using the dot product. The direction is measured by the $z$ component of the exterior product ($\wedge $) between the direction vector  to the target point  ${\vec P}_i$ and the aircraft speed ${\vec V}_i$. With ${\vec P}_i'=P_{i_x}e_x+P_{i_y}e_y$ and ${\vec V}_i'=V_{i_x}e_x+V_{i_y}e_y$, in order to align the aircraft  to the target point in the $(x,y)$ plane, the new bank angle is,
\begin{equation}
\begin{array}{lcl}\displaystyle
\mu_{i+1}^{hea}&=& \displaystyle - T_{hard}\arccos{{\vec P}_i' .{\vec V}_i'\over ||{\vec P}_i || \times || {\vec V}_i ||}
\hbox{Sign} (({\vec P}_i \wedge {\vec V}_i)_z)
\\[10pt] \displaystyle
&=&\displaystyle - T_{hard}\arccos{\left(\frac{P_{i_x}V_{i_x}+P_{i_y}V_{i_y}}{\sqrt{(P_{i_x}^2+P_{i_y}^2)(V_{i_x}^2+V_{i_y}^2)}}\right)}
\hbox{Sign}\left[ P_{i_x}V_{i_y}-P_{i_y} V_{i_x}\right]
\end{array}
\label{bank}
\end{equation}
where, we have introduced a new constant $T_{hard}\in[0,1]$. The higher this constant, the faster the vehicle will turn for the same angular deviation. 

We impose now a security threshold in the bank angle, $\mu_{max}$.  A typical value for  the maximum bank angle is $\mu_{max}=\pm 70^o$.
Therefore, the new control bank angle is,
\begin{equation}
\mu_{i+1}=\min\{|\mu^{hea}_{i+1}|,|\mu_{max}|\}.\hbox{Sign}(\mu^{hea}_{i+1})\, .
\label{con2}
\end{equation}


\end{description}

\section{Simulations}\label{conclu}\label{sec:5}

In the previous section, we have described a control mechanism in order to guide a glider to a target. At each time step, the algorithm determines the shortest path to the target and determines the unique values of the attitude commands of the glider that are compatible with the aerodynamic characteristics of the glider. We now test this algorithm with some numerical simulations. 

\begin{figure}
\begin{centering}
 \includegraphics[width= 0.9\hsize]{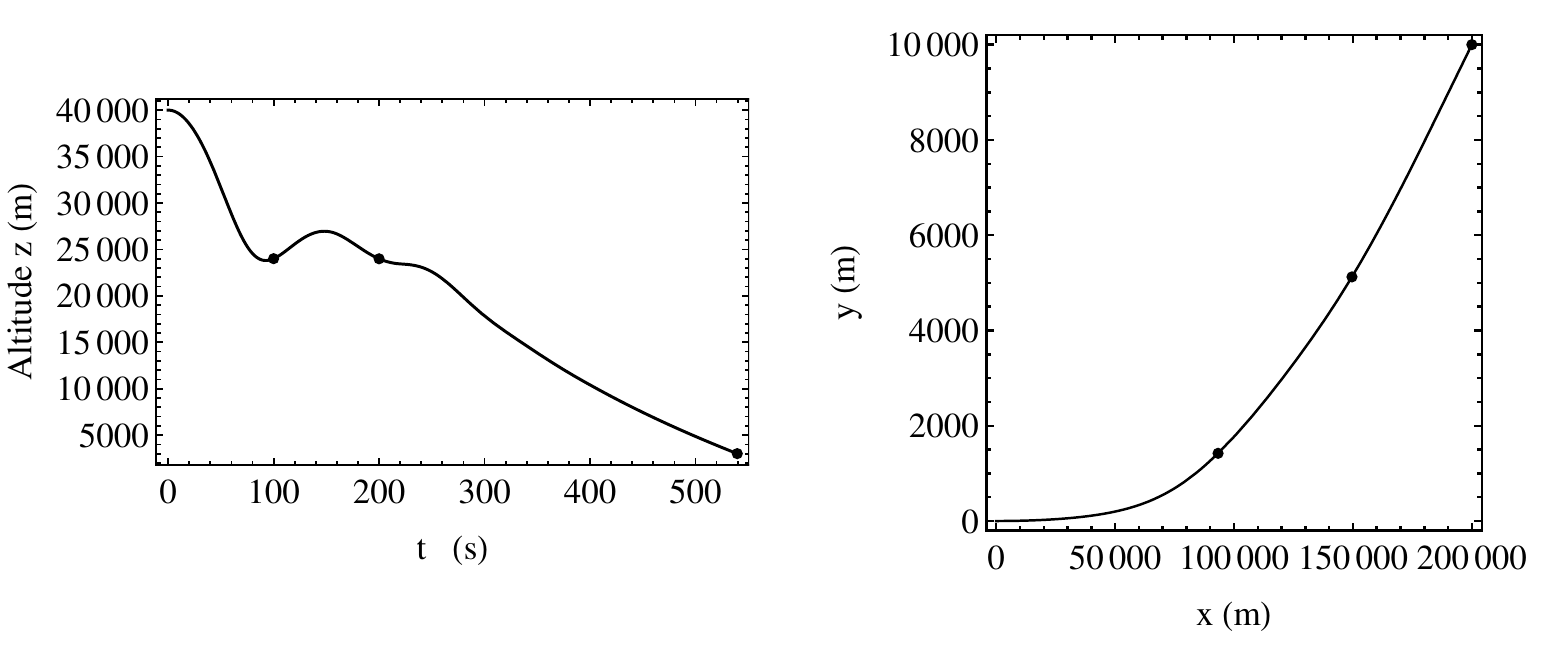}
  \includegraphics[width= 0.9\hsize]{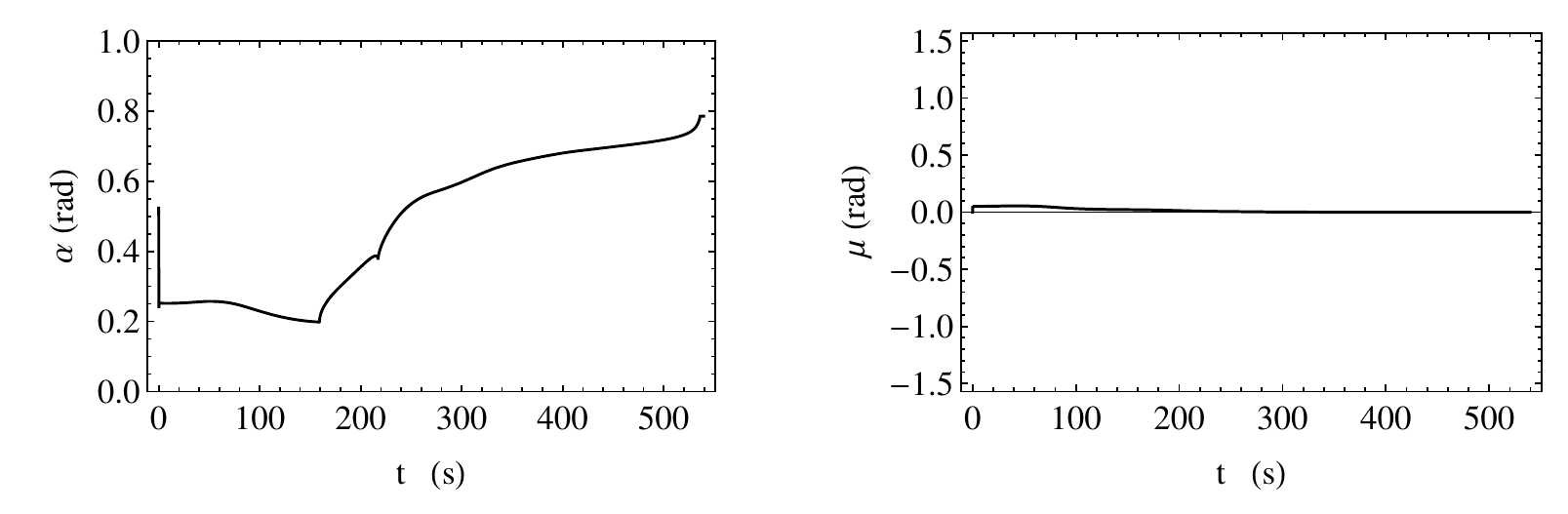}
\caption{Trajectory of the glider in the ambient space and control commands as a function of time. The coordinates of the HAC point target are $(x_f,y_f,z_f)=(200\, 000,10\, 000,3\, 000)$~m. The time of arrival at the HAC is$t=539.6$~s, with a distance error $e_d=14.6$~m and final speed $V_f=0.203$~M. The dots indicate the position of the glider after $100$~s and $200$~s of flight and the HAC position. Positive values of $\mu$ correspond to left turns and negative values of $\mu$ correspond to right turns.}
\label{simu1} 
\end{centering}
\end{figure}

We have taken the glider initial coordinates $(x_0,y_0,z_0)=(0,0,40000)$~m, $V_0=1000$~m/s, $\gamma_0=0$, $\chi_0=0$, $\mu_0=0^o$, $\alpha_0=30^o$, $T_{con}=0.1$ and $T_{hard}=1.0$, and we calculated the trajectories of the glider by numerically intreating equations (\ref{equations2}) with a fourth order Runge-Kutta integration method. 

\begin{figure}
\begin{centering}
 \includegraphics[width= 0.9\hsize]{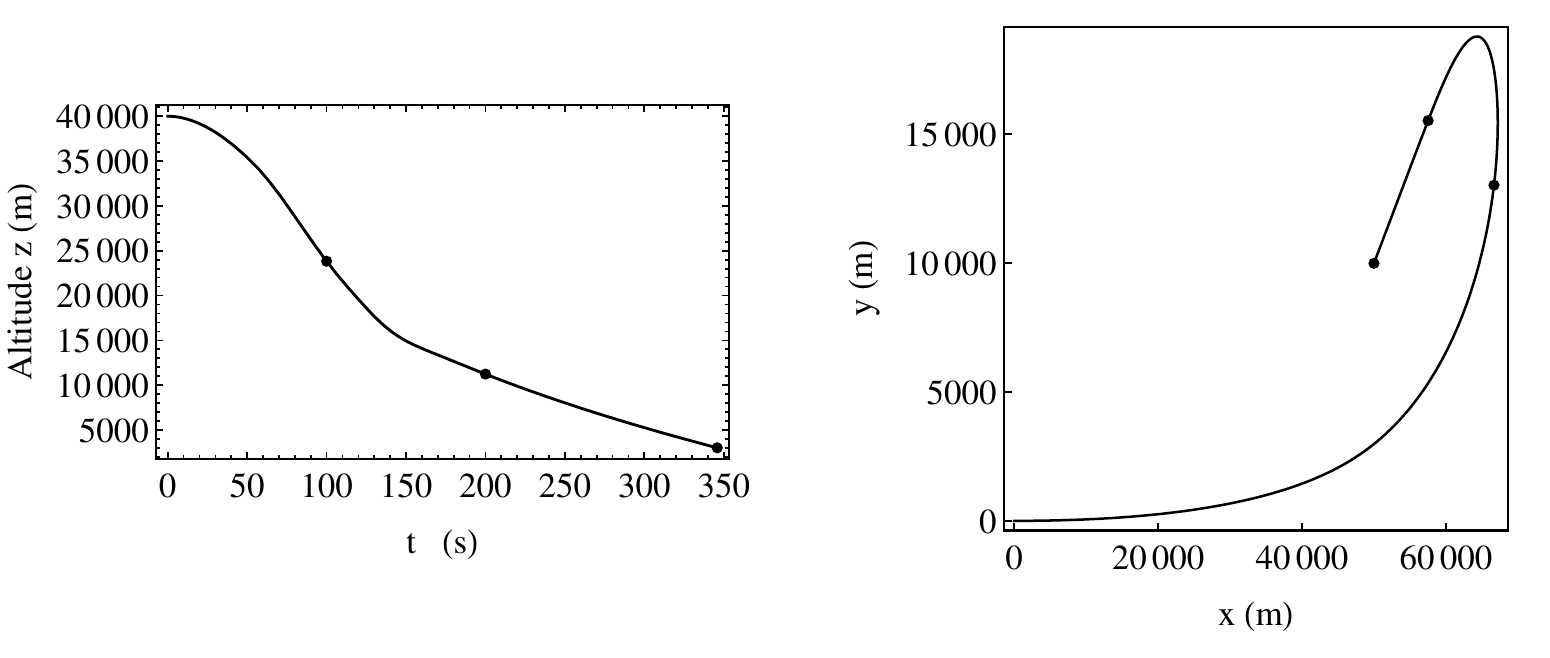}
  \includegraphics[width= 0.9\hsize]{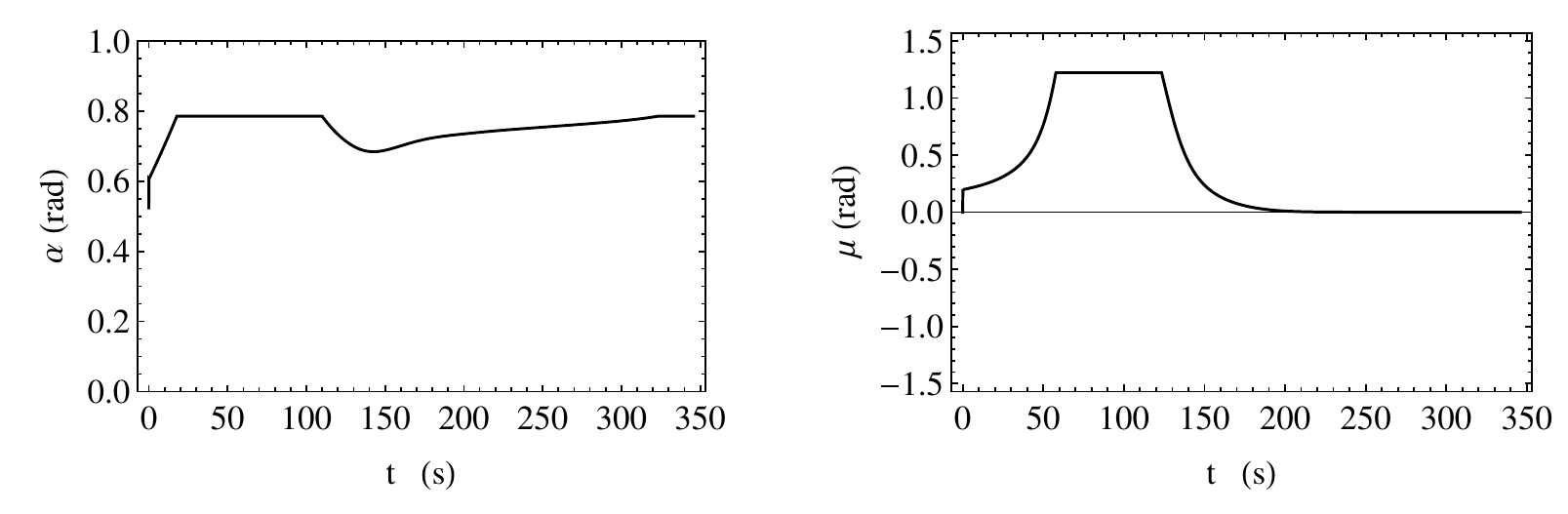}
\caption{Trajectory of the glider in the ambient space and control commands as a function of time. The coordinates of the HAC point target are $(x_f,y_f,z_f)=(50\,000,10\,000,3\,000)$~m. The time of arrival at the HAC  is$t=345.9$~s, with a distance error $e_d=23.1$~m and  final speed $V_f=0.205$~M. The dots indicate the position of the glider after $100$~s and $200$~s of flight and the HAC position.}
\label{simu2} 
\end{centering}
\end{figure}

\begin{figure}
\begin{centering}
 \includegraphics[width= 0.8\hsize]{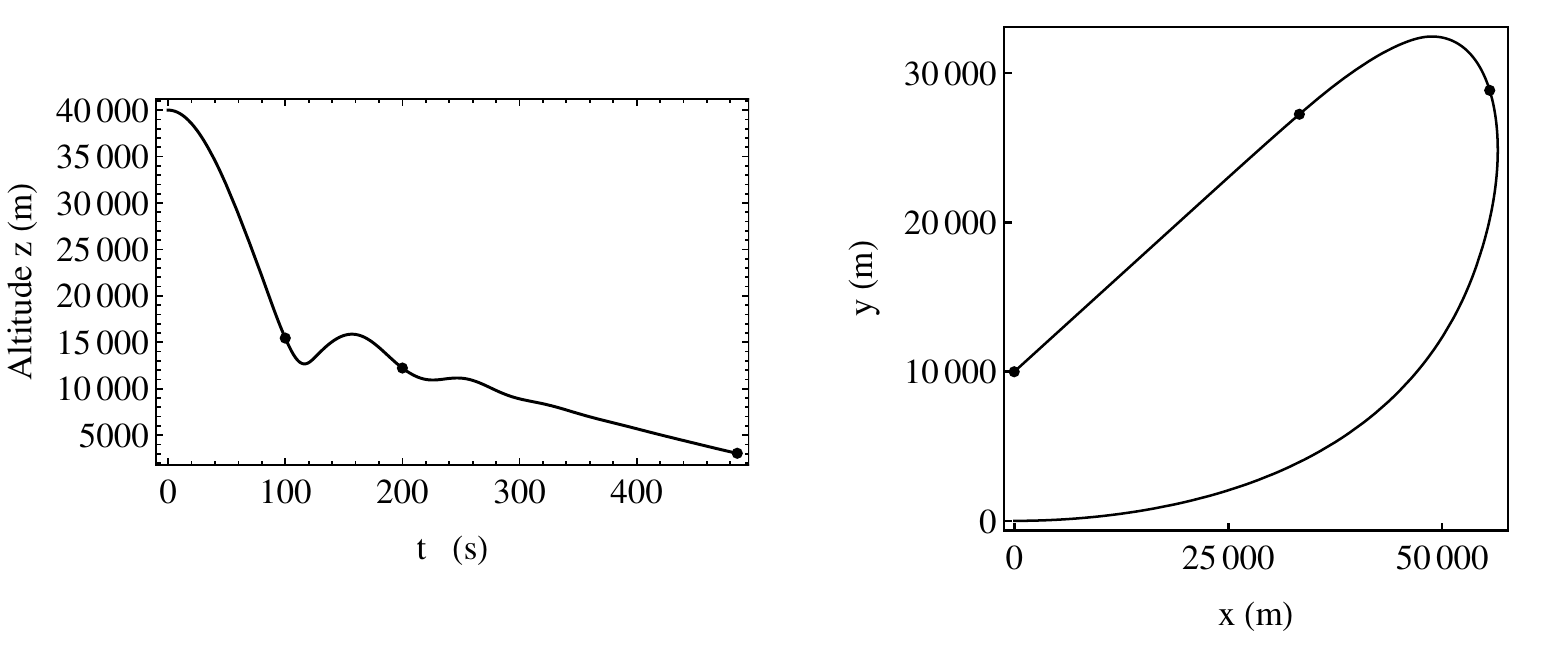}
  \includegraphics[width= 0.8\hsize]{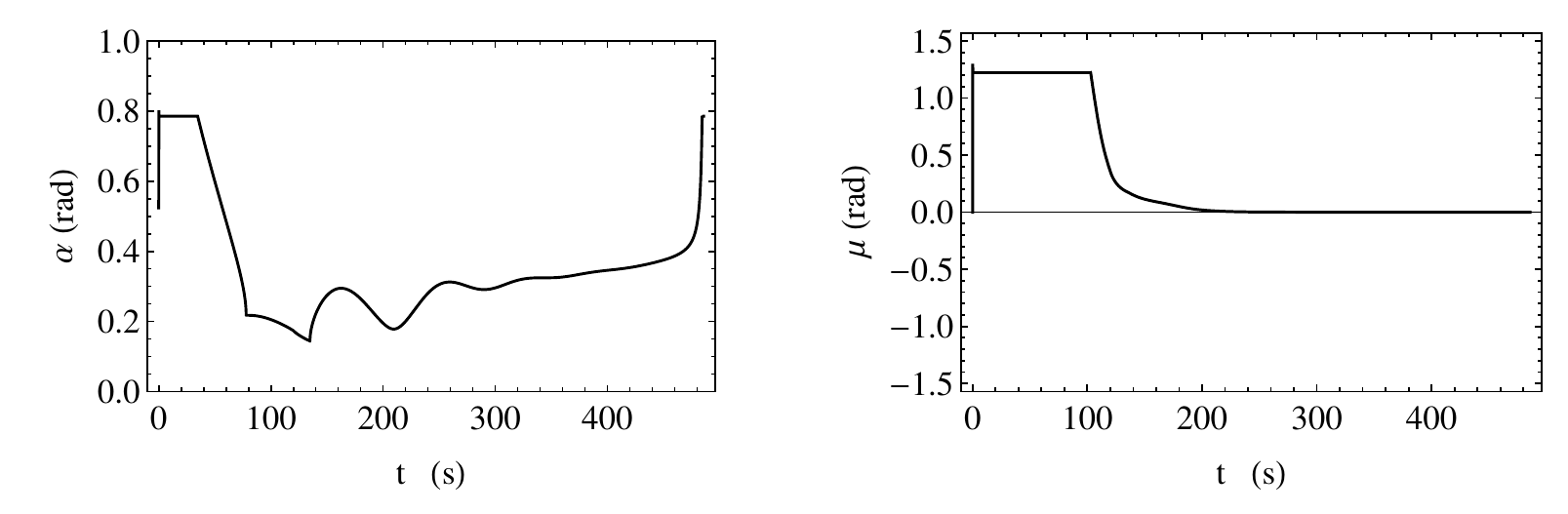}
\caption{Trajectory of the glider in the ambient space and control commands as a function of time. The coordinates of the HAC point target are $(x_f,y_f,z_f)=(0,10\,000,3\,000)$~m. The time of arrival at the HAC  $t=485.9$~s, with a distance error $e_d=51.3$~m and   final speed $V_f=0.200$~M. The dots indicate the position of the glider after $100$~s and $200$~s of flight and the HAC position.}
\label{simu3} 
\end{centering}
\end{figure}

The goal was to reach some target point that we have defined as the  centre point of the HAC. We have chosen three different target HAC points with coordinates,
\begin{description}
\item{1)} $(x_f,y_f,z_f)=(200\,000,10\,000,3\,000)$~m (figure~\ref{simu1}).
\item{2)} $(x_f,y_f,z_f)=(50\,000,10\,000,3\,000)$~m (figure~\ref{simu2}).
\item{3)} $(x_f,y_f,z_f)=(0,10\,000,3\,000)$~m (figure~\ref{simu3}).
\end{description}
and we have calculated the controlled trajectories from the same initial point. 
The arrival to the HAC point  occurs when the distance from the glider to the centre of the HAC point attains a minimum.
This distance  error  will be denoted by $e_d$. In figures~\ref{simu1}, \ref{simu2} and \ref{simu3}, we show the glider controlled trajectories as function of time and the sequence of the attack and bank angle values as computed by the command and control algorithm. We have computed the time of arrival at the HAC, the final speed at the HAC ($V_f$) measured in Mach number units, and the distance error $e_d$.

The basic features of this algorithm is to guide the aircraft to the HAC point with  very low distance errors. The choice of the initial conditions has been done  insuring  that the initial energy of the glider is enough to arrive at the target point.
In this  study, we have chosen target points within the maximum range calculated numerically by imposing the condition that the flight is always done with zero bank angle and maximum glide angle.
In this case, the ratio $C_L/C_D$ is maximal and the drag on the glider is minimal. 
 For the initial conditions chosen and the Space Shuttle parameters, the range is of the order of $286$~km.

\begin{figure}[h]
\begin{centering}
 \includegraphics[width= 0.5\hsize]{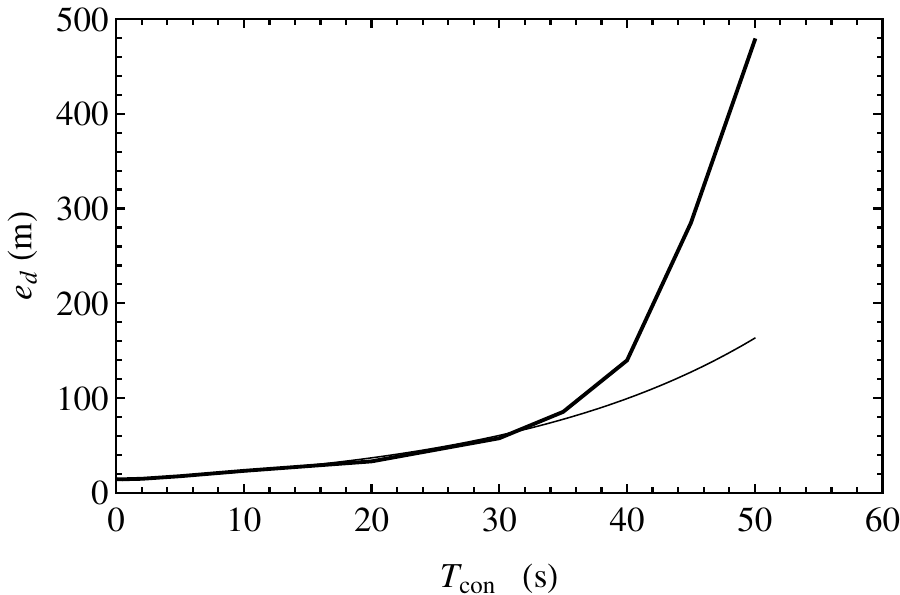}
\caption{Dependence of the distance error on the control time $T_{con}$ for the initial and final conditions of the simulation in figure~\ref{simu1}. For $T_{con}\le 30$, the distance arrow follows the approximate exponential law (\ref{error}), represented by the thin line.}
\label{simu4} 
\end{centering}
\end{figure}

Dynamic aircraft trajectories computed with the algorithm presented here depend on the control time $T_{con}$. For the conditions in figure~\ref{simu1}, we have evaluated the distance error from the centre of the HAC as a function of  $T_{con}$.
For $T_{con}\le 30$, we have found that,
\begin{equation}
 e_{d}=13.7e^{0.049 T_{con}}\, .
 \label{error}
\end{equation}
In figure~\ref{simu4}, we show the dependence of the distance error on the control time $T_{con}$ for the initial and final conditions of the simulation in figure~\ref{simu1}.

\begin{figure}[h]
\begin{centering}
 \includegraphics[width= 0.9\hsize]{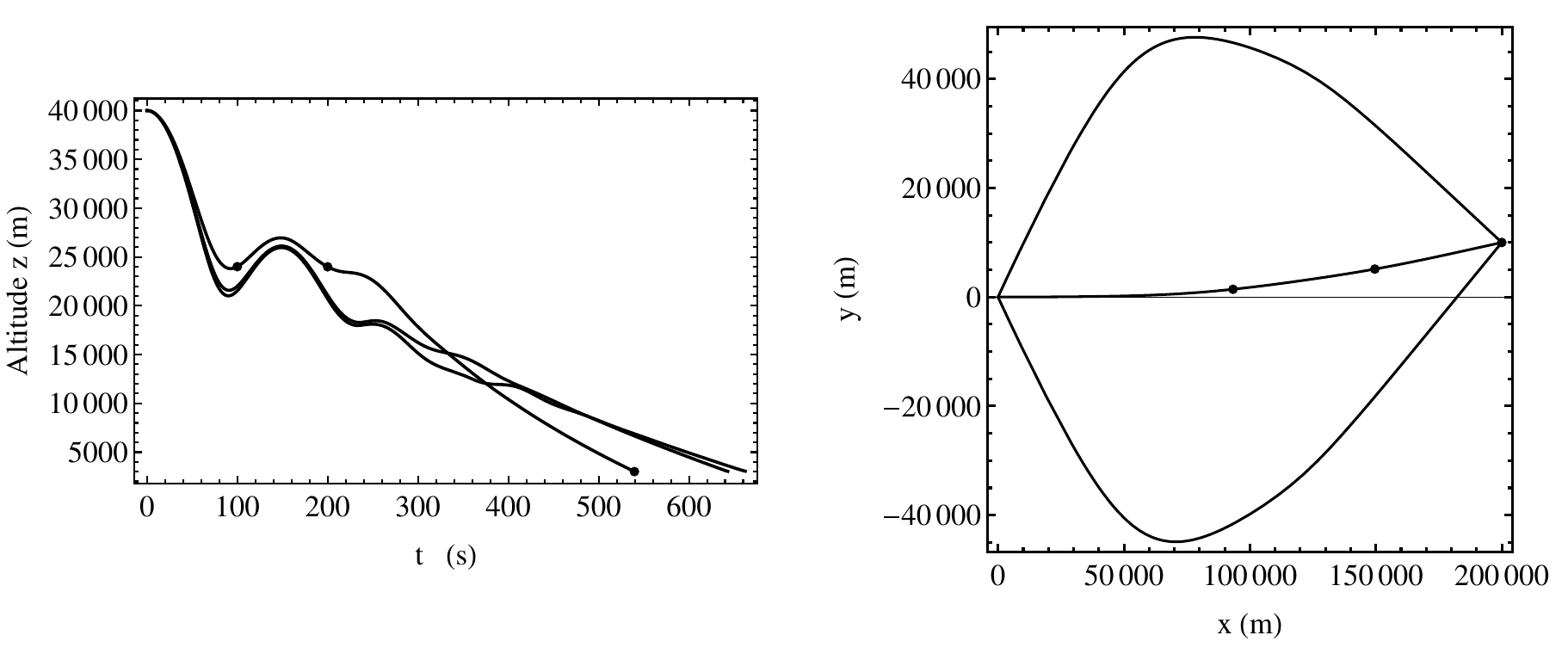}
\caption{Different trajectories calculated with the initial parameter $\chi_0=\pi/4, 0,-\pi/4$. The other parameters are the same as in figure~\ref{simu1}. For these trajectories, the distance errors are $e_d=34.6$~m,  $e_d=14.6$~m and $e_d=52.2$~m, respectively. }
\label{simu5} 
\end{centering}
\end{figure}

We have also tested the dependence of the controlled trajectories as a function of the entry angle $\chi_0$. In  figure~\ref{simu5}, we show the trajectories as in figure~\ref{simu1} but with $\chi_0=\pi/4, 0,-\pi/4$. In this three cases, the distance errors are $e_d=34.6$~m,  $e_d=14.6$~m and $e_d=52.2$~m, respectively. For larger values of angles $\chi_0$, the distance error can be as large as $69$~km ($\chi_0=\pi/2$). 

\section{Conclusions}\label{sec:6}

We have derived a new algorithm for the command and control during the TAEM phase of re-usable space vehicles. 
The algorithm determines locally the shortest path to the target point, compatible with the aerodynamic characteristics of the aircraft. We have tested the ability of the algorithm to guide the Space Shuttle during the TAEM re-entry orbit, proving the feasibility of the algorithm, even using control times of the order of $30$~s. Further refinements of the algorithm are under study \cite{11}.

\begin{acknowledgement}
This work has been developed in the framework of a cooperation with AEVO GmbH (Munich) and
we would like to acknowledge Jo\~ao Graciano
 for suggestions and critical reading of this paper. RD would like to thank IH\'ES, where the final version of the paper has been prepared.
\end{acknowledgement}
\section*{Appendix}
\addcontentsline{toc}{section}{Appendix}
The Earth atmosphere parameters are based on the Ò1976 US Standard Atmosphere ModelÓ. For the first seven layers we have used the formulas described in \cite{1}. In table~\ref{tab2}, we show the parameterisation of the thermodynamic quantities for the Earth atmosphere.

\begin{table}[h!]
\begin{center}

\begin{tabular}{|c|c|c|c|c|}
\hline
Layer & $z_0$ (m) &$T_0$ (K) &$\lambda_0 $(K/m) & $P_0$ (Pa) \\
\hline
1 & $0$  & $288.15$ & $-0.0065$ & $101325.00$  \\
\hline
2 & $11 019$ &  $216.65$ &  $-$ & $22632.10$   \\
\hline
3 & $20 063$ &  $216.65$ &  $0.0010$ & $5474.89$   \\
\hline
4 & $32 162$ &  $228.65$ &  $0.0028$ & $868.02$   \\
\hline
5 & $47 359$ &  $270.65$ &  $-$ & $110.91$   \\
\hline
6 & $51 412$ &  $270.65$ &  $-0.0028$ & $66.94$   \\
\hline
7 & $71 802$ &  $214.65$ &  $-0.0020$ & $3.96$   \\
\hline
\end{tabular}
\end{center} 
\begin{center}
\begin{tabular}{|c|c|c|c|}
\hline
Layer & $T$ (K) & $P$ (Pa) & $\rho$ (kg/m$^3)$\\ [3pt]
\hline   
1 &  $T_0+\lambda_0 (z-z_0)$ & $P_0 (\frac{T_0}{T})^{g(z)M_{air}/(R  \lambda_0)}$ & $\frac{P}{TR_s}  $\\    [3pt]
\hline  
2 &  $T_0$  & $P_0 e^{-g(z)M_{air}(z-z_0)/(R  T)}$   & $\frac{P}{TR_s}  $\\ [3pt]
\hline
3&  $T_0+\lambda_0 (z-z_0)$     & $P_0 (\frac{T_0}{T})^{g(z)M_{air}/(R  \lambda_0)}$ & $\frac{P}{TR_s}  $\\ [3pt]
\hline
4 &  $T_0+\lambda_0 (z-z_0)$    &$P_0 (\frac{T_0}{T})^{g(z)M_{air}/(R  \lambda_0)}$ & $\frac{P}{TR_s}  $\\
\hline
5 &  $T_0$  & $P_0 e^{-gM_{air}(z-z_0)/(R  T)}$   & $\frac{P}{TR_s}  $\\
\hline
6 &  $T_0+\lambda_0 (z-z_0)$     & $P_0 (\frac{T_0}{T})^{g(z)M_{air}/(R  \lambda_0)}$ & $\frac{P}{TR_s}  $\\
\hline
7 &  $T_0+\lambda_0 (z-z_0)$   & $P_0 (\frac{T_0}{T})^{g(z)M_{air}/(R  \lambda_0)}$ & $\frac{P}{TR_s}  $\\
\hline
\end{tabular}

\end{center}
\caption{Characteristic parameters for the lower layers of the atmosphere.   $z_0$ is the lower altitude  of the layer, $R = 8.31432$~J/(mol~kg) and $R_s=287.04$~J/(kg~K) are gas constants, $M_{air}=0.0289644$~kg/mol, $g(z)=g_0(R_E/(R_E+z))^2$ is the gravity acceleration, $g_0=9.80665$~m/s$^2$ is the standard gravitational acceleration constant  and $R_E=6.371\times10^6$~m is the Earth mean radius.}
\label{tab2}
\end{table}


\end{document}